\documentstyle[aps,amssymb,preprint,prb,tighten,floats,
amsfonts,times
]{revtex}

\input{definitionen.tex}
\draft
\begin{document}

\draft
 
\title{A rigorous path integral for quantum spin using flat-space\\ Wiener regularization}
\author{Bernhard Bodmann}
\address{Department of Mathematics,
      University of Florida, 358 Little Hall,
      Gainesville, Florida 32611, USA}
\author{Hajo Leschke and Simone Warzel}
\address{Institut f\"ur Theoretische Physik,
      Universit\"at Erlangen-N\"urnberg,
      Staudtstr. 7, D-91058 Erlangen, Germany}
%
%
\maketitle
\begin{abstract}%
Adapting ideas of Daubechies and Klauder [J.\ Math.\ Phys.\ {\bf 26}, 2239 (1985)]
we derive a rigorous continuum path-integral formula for the
semigroup generated by a spin Hamiltonian.
More precisely, we use spin coherent vectors parametrized by complex
numbers to relate the coherent representation of this  
semigroup to a suitable Schr\"odinger semigroup on the Hilbert space 
${\mathrm L}^2({\mathbb R}^2)$ of Lebesgue square-integrable functions on the Euclidean
plane ${\mathbb R}^2$.
The path-integral formula emerges from 
the standard Feynman-Kac-It\^o formula for
the Schr\"odinger semigroup in the ultradiffusive limit of the underlying
Brownian bridge on ${\mathbb R}^2$. 
In a similar vein, a path-integral formula can be constructed for the coherent
representation of the unitary  
time evolution generated by the spin Hamiltonian.
\end{abstract}
\pacs{PACS numbers: 02.50.Ey, 75.10.Jm\\[.5cm]
  {\footnotesize \it Appeared in:} J. Math. Phys. {\bf 40} 2549--2559 (1999)}

\section{Introduction}
Even 50 years after the appearance of Feynman's celebrated paper 
\cite{Fey48}
that introduced
the path-integral formalism \cite{FeHi65,Sim79,Sch80,Kle95,Roe94}
into quantum theory
in a heuristic but convincing manner, there is 
no general consensus on how to treat a quantum spin within this 
framework.
To the best of our knowledge, among the various approaches over the years, see, for example, Refs. 
\onlinecite{Mar59a,Mar59b,Kla60,Kla79,KuSu80,DaKl85,Sch68,IJR98,NiRo88,Joh89,FrSt88,Sto89,JaKu91,CDGR97,AnJo82,Lem96}, 
the only rigorous
expression for the dynamics of a quantum spin in terms of an integral over continuous
paths is due to Daubechies and Klauder.\cite{DaKl85}
These authors were able to write the coherent representation of the unitary time-evolution 
operator of a spin with a definite quantum number
as a Wiener-regularized path integral, more precisely, as the ultradiffusive limit of a 
well-defined integral over spherical Brownian-motion paths.

The main goal of the present paper is to show that one may equally well 
perform the Wiener regularization by employing planar Brownian motion. 
In this way also a closer contact to symbolic continuum path-integral
formulas widely discussed in the recent literature 
\cite{Kur92,FKSF95,FKNS95,Koc95,EMNP96,ShTa98} is established.
One may hope that the wealth of analytical tools associated with the flat-space
Wiener measure helps clarifying some subtle points there.

%
%
\section{Basic Definitions, Result, and Comments} \label{defres}
%
%
%
%
We consider a single spin with fixed {\it quantum number} 
$j \in \left\{0, 1/2, 1, 3/2, \dots \right\}$, that is, 
using physical units where Planck's constant $ 2 \pi \hbar$ equals $2 \pi$,
        \begin{equation}
                \half \left( {\cal J_+} {\cal J_-} + {\cal J_-} {\cal J_+} \right) + 
                {{\cal J}_3}^2 = j(j+1) \one \, .
        \end{equation}
The {\it spin operators} ${\cal J_+}, {\cal J_-}$ and $ {{\cal J}_3}$  
obey the usual angular-momentum commutation relations
$
   {\cal J_+} {\cal J_-} - {\cal J_-} {\cal J_+} = 2 {{\cal J}_3} $, $ {{\cal J}_3} {\cal J_\pm} - {\cal J_\pm} {{\cal J}_3} = \pm {\cal J_\pm} 
$
and are viewed as acting on the $(2j+1)$-dimensional complex Hilbert 
space ${\mathbb{C}}^{2j+1}$. Its standard scalar product
is denoted as $\braket{\bcdot}{\bcdot}$ and, by convention, antilinear in  the first argument.
The unit operator on ${\mathbb{C}}^{2j+1}$ is denoted by $\one$.
%
%

Non-normalized so-called {\it coherent vectors}\cite{Kla60,Rad71} in this Hilbert space,
        \begin{equation}
                \ket{z} :=  g(z) \, e^{z {\cal J_+}} \ket{j,-j} \, , \qquad z \in {\mathbb{C}}\, ,
        \end{equation}
are parametrized by complex numbers $z$. Henceforth,  $z^*$ will refer to
their complex conjugates, $z_1:=(z+z^*)/2$ and $z_2:= (z-z^*)/2i$
to their real and imaginary parts, and we write $f^*(z):=(f(z))^*$ for the values 
of complex-conjugated functions $f^*$. 
For later notational convenience the strictly positive prefactor is taken as 
        \begin{equation}\label{weight}
                g(z):=\left(\frac{2j+1}{\pi}\right)^{1/2}
                        \left(1 + |z|^2 \right)^{-j-1},
        \end{equation}
and a normalized {\it spin-down vector} $\ket{j,-j} \in {\mathbb{C}}^{2j+1}$, 
obeying
$
        {\cal J_-} \ket{j,-j} = 0
$ and
$
        \braket{j,-j}{j,-j} = 1 \, ,
$
serves as the reference vector.
Every vector $\ket{\psi} \in {\mathbb{C}}^{2j+1}$ is
characterized by its so-called 
{\it coherent representation} $\braket{z}{\psi}$, a function of the form 
$g(z)$ times a polynomial in $z^*$ of maximal degree $2j$.
The scalar product of two coherent vectors
$\langle z|z'\rangle = g(z)g(z')(1 + z^* z')^{2j}$
is an example.
Given an arbitrary operator $\cal B$ on ${\mathbb{C}}^{2j+1}$, the scalar product
$\bra{z}{\cal B}\ket{z'}$ of $\ket{z}$ and ${\cal B}\ket{z'}$ is called the 
{\it coherent representation} of $\cal B$. The mapping 
$(z,z') \mapsto \bra{z}{\cal B}\ket{z'}$ is continuous, because $z \mapsto \ket{z}$
is continuous, every operator $\cal B$ on ${\mathbb{C}}^{2j+1}$ 
is bounded, and the scalar product 
$(\ket{\varphi}, \ket{\psi}) \mapsto \langle\varphi|\psi\rangle$ is continuous.
An example is $\bra{z}\, e^{2\lambda {\cal J}_3 } \ket{z'} = g(z) g(z')
\left( e^{-\lambda} + z^* z' e^\lambda \right)^{2j}$,
$\lambda \in \mathbb{C}\,$.

In what follows, it is a comforting fact that whatever the 
{\it spin Hamiltonian} ${\mathcal H}$ may be -- 
given as a (self-adjoint) operator on ${\mathbb{C}}^{2j+1}$ --  it is polynomial
in the spin operators ${\cal J_+}, {\cal J_-}$, and ${{\cal J}_3}$, and  
it is always possible to write it in {\it pseudodiagonal form},
        \begin{equation} \label{Hdef}
                {\mathcal H} = \int_{\mathbb{C}} \! d^2 z \, h(z) \, \ket{z} \bra{z} \; .
        \end{equation}
Here the (real-valued) 
function $h$ on ${\mathbb{C}} \cong  \mathbb{R} \times \mathbb{R}$
may be chosen bounded and continuous,\cite{KlSk85,Per86,ZFG90} 
the operator $\ket{z} \bra{z}/\langle z|z\rangle$ denotes the orthogonal projection onto the 
one-dimensional subspace spanned by $\ket{z} \in {\mathbb{C}}^{2j+1}$,  and 
$d^2z := dz_1 dz_2$ is the two-dimensional Lebesgue measure 
on the Euclidean plane ${\mathbb{R}} \times {\mathbb{R}} =: {\mathbb{R}}^2$. 
Following  Ref. \onlinecite{Ber72a}, we call $h$ a {\it contravariant symbol} of $\cal H$, 
elsewhere called an upper \cite{Sim80} or lower \cite{Kla98} symbol.
In particular, the unit operator $\one$ has the constant $1$ as a contravariant symbol.
In this sense, the coherent vectors are {\it unity-resolving} and hence
(over-)complete. 
Other examples for contravariant symbols
are listed in Table \ref{tabelle}; confer  Ref.  \onlinecite{Lie73}.
\begin{center}
\begin{table}[hbtp] 
\begin{tabular}{c c c || c c }
Operator & Contravariant symbol & \qquad & Operator & Contravariant symbol\\ \hline 
&&&&\\[-0.4cm] 
${\cal J_+}$ &  $2(j+1) \frac{z^*}{1+|z|^2}$ & & ${\cal J_+} {\cal J_-}$  &  $-2(j+1) \, \frac{1 - 2(j+1)\abs{z}^2}{(1 + \abs{z}^2)^2}$ \\[0.3cm]
${\cal J_-}$  &  $2(j+1) \frac{z}{1+|z|^2}$  & & ${\cal J_-}{\cal J_+}$  &  $2(j+1) \, \frac{2(j+1)\abs{z}^2 - \abs{z}^4}{(1 + \abs{z}^2)^2}$ \\[0.3cm]
${{\cal J}_3}$ &  $ -(j+1) \frac{1-|z|^2}{1+|z|^2}$ & & ${{\cal J}_3}^2$   &  $ (j+1)(j+\frac{3}{2}) \, \left(\frac{1 - \abs{z}^2}{1 + \abs{z}^2}\right)^2 - \frac{j+1}{2}$ \\[-0.4cm]
&&&& 
\end{tabular}
\caption{Contravariant symbols for selected operators on ${{\mathbb{C}}^{2j+1}}$, which are bounded and continuous. \label{tabelle}}
\end{table}
\end{center}
%
%
After these preparations we are able to state the main result of the 
present paper, namely, a rigorous expression for the {\it spin semigroup} 
$\bigl\{e^{-t {\mathcal H}}\bigr\}_{t \ge 0}$ as the {\it ultradiffusive limit} 
of a Wiener type of integral over Brownian-motion paths 
$\left\{s \mapsto b(s)=b_1(s)+ib_2(s)\right\}_{s \geq 0}$
on the complex plane $\mathbb{C} \cong \mathbb{R} \times \mathbb{R}$ . More 
precisely, the coherent representation of
$\ep{-t {\mathcal H}}$ may, for all $z,z' \in {\mathbb{C}}$ and $t>0$, be written as    
        \begin{eqnarray} \label{result}
                \bra{z} \ep{- t {\mathcal H}} \ket{z'} & = &
                 \lim_{\nu \to  \infty} \dmut 
                \exp\left\{4(j+1)\nu   \int_0^t 
                        \! \frac{ds}{(1 + |b(s)|^2)^2}\right\} \nonumber \\ 
                && \qquad \qquad \times \exp\left\{(j+1) \int_0^t  \! ds \, 
                \frac{\dot{b}(s) b^*(s) - \dot{b}^*(s) b(s)}{1+|b(s)|^2}
                 -  \int_0^t \! ds \, h(b(s))\right\} \; .
        \end{eqnarray}
Here for given $z$, $z' \in \mathbb{C}$, $t>0$, and $\nu>0$ the path integration
is defined by
        \begin{equation}
                \dmut \big(\bcdot\big) :=
                \frac{1}{4 \pi t \nu} \,  
                \ep{- |z-z'|^2/4t \nu} \,
                {\mathbb{E}}( \bcdot ) \; , 
        \end{equation}
where $\mathbb{E}(\bcdot)$ indicates the probabilistic expectation with 
respect to the {\it two-dimensional Brownian bridge},
with diffusion constant $\nu$ starting in $z=b(0)$ and arriving at $z'=b(t)$
a time $t$ later.\cite{Sim79,Roe94,RoWi87,Pro95,ReYo94} 
As a Gaussian stochastic process with continuous paths on 
$\mathbb{C} \cong \mathbb{R} \times \mathbb{R}$
the Brownian bridge, in its turn, is uniquely determined by its mean,
        \begin{equation}
                {\mathbb{E}}\left(b(s)\right) = z + (z'-z) \frac{s}{t}
                \, , \mkern230mu s \in [0,t] \qquad
        \end{equation}
and covariances,
        \begin{eqnarray}
                {\mathbb{E}}\left(b^*(r) b(s)\right) - 
                {\mathbb{E}}\left(b^*(r)\right)\, {\mathbb{E}}\left(b(s)\right)
                & = & 4\nu \left( \min\{r,s\} - \frac{rs}{t} \right)\, , \\
                {\mathbb{E}}\left(b(r) b(s)\right)  - 
                {\mathbb{E}}\left(b(r)\right)\, {\mathbb{E}}\left(b(s)\right)
                & = & 0  
                \, , \mkern200mu  r,s \in [0,t] \, . \qquad
        \end{eqnarray}
The second integral in the exponent on the right-hand side of \eq{result} 
is a purely imaginary stochastic (line) integral, \cite{RoWi87,Pro95,ReYo94} 
which is understood 
in the sense of Fisk and
Stratonovich and to which one is therefore allowed to apply 
the rules of ordinary calculus,\cite{RoWiTheorem} although the 
time derivative {$\dot{b}\;$} does not exist.\\\

Several comments apply.
\begin{enumerate}
\item 
By the It\^o formula \cite{Sim79,RoWi87,Pro95,ReYo94}
it can be seen that the stochastic integral in \eq{result} may equally well be interpreted as a 
stochastic integral in the sense of It\^o. Moreover, using the
It\^o formula in a different way, the sum of this integral and the
first (Lebesgue) integral
in the exponent of the right-hand side of \eq{result}
can be converted \cite{ReYoTheorem} according to 
   \begin{equation} \label{umrech}
                4\nu \int_0^t \! \frac{ds}{(1 + |b(s)|^2)^2}    
                 + \int_0^t \! ds \,  
                \frac{\dot{b}(s) b^*(s) - \dot{b}^*(s) b(s)}{1+|b(s)|^2}
                =  \ln 
                \left(\frac{1 + |b(t)|^2}{1 + |b(0)|^2}\right) - 
                2 \int_0^t  \! 
                \frac{db^*(s) b(s)}{1+|b(s)|^2}  
                \, .
        \end{equation} 
Here the complex stochastic integral 
$\int_0^t  \! db^*(s) b(s)/\left[1+|b(s)|^2\right]$
has to be understood in the sense of It\^o. It contains the
only $\nu$-dependence of the right-hand side.
By using \eq{umrech} in the path integrand in \eq{result} the logarithmic term
results in the prefactor $\left[(1+|z'|^2)/(1+|z|^2)\right]^{j+1}=g(z)/g(z')$. 
\item 
The stochastic integral in \eq{result} is of kinematical origin and
reflects the symplectic structure,
which renders the complex plane a phase space for the so-called
classical spin;\cite{Ber74a,Ber74b,Per86} also see the concluding remarks.
\item
If one wants to use \eq{result} to express the trace $\int_{\mathbb C} d^2 z
\bra{z} \ep{-t {\mathcal H}} \ket{z} $ of $\ep{-t {\mathcal H}}$ as a path integral,
one should resist the temptation to interchange the integration with respect to $z$ with
 the ultradiffusive limit $\nu \to \infty$, because the resulting prelimit expression
 would be infinite.
\item
Instead of taking the ultradiffusive limit, 
one may perform the regularization also
by a {\it long-time limit}, in the sense that
        \begin{eqnarray}  \label{res}
                \bra{z} \ep{- t {\mathcal H}} \ket{z'} & = &
                \lim_{u \to  \infty}  \dmuu    
                \exp\left\{4(j+1)\nu   \int_0^u 
                        \! \frac{ds}{(1 + |b(s)|^2)^2}\right\} \nonumber \\ 
                && \quad \times \exp\left\{(j+1) \int_0^u ds \,
                \frac{\dot{b}(s) b^*(s) - \dot{b}^*(s) b(s)}{1+|b(s)|^2}
                 -  \frac{t}{u} \int_0^u \! ds \, h(b(s))\right\} \; .
        \end{eqnarray}
This formula can be deduced from \eq{result} by suitably scaling the Brownian bridge, 
holds for all $\nu > 0$, 
and, in contrast to \eq{result}, 
makes sense as it stands even for $t \leq 0$, 
hence for all $t \in \mathbb{R}$. One should notice that the time-parameter set of the
Brownian bridge used in (\ref{res}) is the closed interval $[0,u]$ and
not $[0,t]$.
\item
Replacing $h$ by $ih$ in \eq{result} or \eq{res} yields analogous 
expressions for the coherent representation of the (unitary) 
{\it spin time-evolution}
operator $\ep{- i t {\mathcal H}}$. A rigorous justification
relies on the boundedness and continuity of $h$ and requires 
extending the subsequent proof by showing analyticity of both sides of \eq{result}
in  a coupling parameter $\lambda \in {\mathbb{C}}$ multiplying $h$. 
The left-hand side and the prelimit 
expression in \eq{result} are easily seen to be analytic in $\lambda$. 
Analyticity in $\lambda$ in
the limit $\nu \rightarrow \infty$ is then proved with the help of 
an equation analogous to \eq{semi} and uniform convergence in $\nu > 2 \nu_0 > 0$ 
of the perturbation
series in $\lambda$ of the relevant operator and functions there.  
\item
The flat-space Wiener-regularized path-integral expression \eq{result}
for the spin semigroup is an alternative to a result 
first given and proved in  Ref.  \onlinecite{DaKl85}. 
There the authors integrate over Brownian-motion paths
on the unit-sphere in the three-dimensional Euclidean space ${\mathbb R}^3$ to 
obtain the coherent representation of 
$\ep{- i t {\mathcal H}}$. Unlike in Ref. \onlinecite{DaKl85},
the regularizing path measure
$\Idmut \exp\{{4(j+1) \nu \int_0^t ds (1 + \abs{b(s)}^2)^{-2}}\}$
used in \eq{result} is not invariant under
the full special unitary group $SU(2)$
when the latter is realized by suitable M\"obius transformations on the
(extended) complex plane.
Yet in the limit $\nu \rightarrow \infty$
all symmetries of a given spin Hamiltonian are restored.
Contrary to what one might expect, Eq. \eq{result} cannot be obtained 
from the corresponding result in  Ref.  \onlinecite{DaKl85}
merely by stereographically
projecting the paths from the sphere onto the (extended) plane. 
Nevertheless, the proof given in the next section shows that the key
ideas behind 
both constructions are the same; also see the concluding remarks.
\item
So far we have considered a fixed spin quantum number $j$.
In order to make contact with the
Wiener-regularized path-integral expression associated with 
a canonical degree of freedom,
also proved in Ref. \onlinecite{DaKl85},
one has to contract \cite{ACGT72,Gil74}
the algebra of $SU(2)$ to the 
Heisenberg-Weyl algebra by taking the {\it high-spin limit} $j \to \infty$.
More explicitly, in the given (polynomial) spin Hamiltonian
$\cal H$ on ${{\mathbb{C}}^{2j+1}}$, one has to replace ${\cal J_+}, {\cal J_-}$, and ${{\cal J}_3}$
by ${\cal J_+}/\sqrt{2j}, {\cal J_-}/\sqrt{2j}$, and ${{\cal J}_3}+j \one$, respectively. 
If ${\cal H}_j = \int_{\mathbb{C}} d^2z \, h_j(z) \ket{z}\bra{z}$ 
denotes the resulting operator, one then finds the relation 
\begin{equation} \label{contract}
   \lim_{j \to \infty} \frac{\pi}{2j} \bra{z/\sqrt{2j}} \,\ep{- t {\cal H}_j}\, \ket{ z'/\sqrt{2j}}
     = \langle\!\langle z| \,\ep{- t {\mathsf H}}\, |z'\rangle\!\rangle
\end{equation}
where $|z\rangle\!\rangle \in {\mathrm L}^2(\mathbb R)$ is a normalized 
canonical coherent vector \cite{KlSk85,Per86,ZFG90}
and the Hamiltonian $\mathsf H$ on ${\mathrm L}^2(\mathbb R)$, the Hilbert space of 
Lebesgue square-integrable complex-valued functions 
on the real line $\mathbb R$, is defined by
\begin{equation}
   {\mathsf H} := \int_{\mathbb{C}} \! \frac{d^2z}{\pi}\,
        {\mathsf h} (z) |z\rangle\!\rangle \langle\!\langle z| 
\quad \mbox{with} \quad {\mathsf h}(z) := \lim_{j \to \infty} \, h_j(z/\sqrt{2j})\, .
\end{equation}
By using \eq{result} for the prelimit expression in \eq{contract},
suitably rescaling the Brownian bridge,
and interchanging the order of the limits $j \to \infty$ and 
$\nu \to \infty$, one arrives at the path-integral formula
\begin{eqnarray}
\langle\!\langle z| \,\ep{- t {\mathsf H}}\, |z'\rangle\!\rangle &=&
 \pi \lim_{\nu \to  \infty} \ep{2t\nu} \dmut  
                \exp\left\{\half \int_0^t ds 
                \left[\dot{b}(s) b^*(s) - \dot{b}^*(s) b(s)\right]\right\}
                        \nonumber \\ 
                                && \qquad \qquad \qquad \qquad \times 
                \exp\left\{- \int_0^t \! ds \, {\mathsf h}(b(s))\right\} \; ,
\label{contraction}
\end{eqnarray}
in agreement with Eq. (1.3) in  Ref.  \onlinecite{DaKl85};
also see Refs. \onlinecite{DaKl86} and \onlinecite{Kla98}.
Formula (\ref{contraction}) can be shown to hold
not only for the polynomial Hamiltonians $\mathsf H$ resulting from the
contraction, but for a wider class of operators whose conditions are
stated in Theorem 2.4 of Ref. \onlinecite{DaKl85}.
\item
With regard to some of the symbolic path-integral expressions for spin systems
frequently encountered
in the literature, see for example, Refs. 
\onlinecite{Kla79,KuSu80}, and
\onlinecite{Kur92,FKSF95,FKNS95,Koc95,EMNP96,ShTa98}, it might be
illuminating to recognize certain formal similarities between these expressions and the above result
\eq{result}. While the kinematical and dynamical terms
in the exponents of all the corresponding path integrands 
look essentially the same, only the above result is based on a
genuine path measure, namely,  
$\Idmut \exp\{{4(j+1) \nu \int_0^t ds (1 + \abs{b(s)}^2)^{-2}}\}$, but requires
taking the limit $\nu \to \infty$. Here, the Wiener type of measure 
$\Idmut$ is often symbolically written
as $\delta^2 b \, \delta(b(0)-z)
\delta(b(t) - z') \exp\{ - (1/4 \nu) \int_0^t ds | \dot b(s)|^2\}$,
or similarly. In any case, the necessity to regularize 
by some ultradiffusive limit was
observed several times also in non-rigorous
works.\cite{Kla79,FrSt88,Sto89,JaKu91,EMNP96}
\end{enumerate}

%
%
%
\section{Proof} \label{deriv}
%
%
The proof of \eq{result}
consists of three major steps, adapting key ideas of  Ref.  \onlinecite{DaKl85}. 
First, the spin Hilbert space ${{\mathbb{C}}^{2j+1}}$ is embedded into ${\mathrm L}^2({\mathbb{C}})$, 
the Hilbert space of Lebesgue 
square-integrable complex-valued functions on ${\mathbb{C}}$. 
Next, it is identified with
the $(2j+1)$-dimensional ground-state eigenspace of a suitable Schr\"odinger operator $R$ acting 
on ${\mathrm L}^2({\mathbb{C}})$.
Then the spin semigroup, now realized on ${\mathrm L}^2({\mathbb{C}})$, is shown to be the  limit 
$\nu \to \infty$ of a Schr\"odinger semigroup generated by a suitably 
perturbed $\nu R$.
Rewriting this Schr\"odinger semigroup with the help of the standard
Feynman-Kac-It\^o path-integral formula finally gives \eq{result}.
%
%
%
\subsection{The embedding of the spin Hilbert space}\label{einbetten}
%
%
The embedding of the spin Hilbert space ${{\mathbb{C}}^{2j+1}}$ into the infinite-dimensional 
Hilbert space ${\mathrm L}^2({\mathbb{C}})$,
equipped with the standard scalar product 
$(\varphi|\psi):=\int_{\mathbb{C}} d^2z \, \varphi^*(z)\psi(z)$,
is accomplished by interpreting the coherent representation as a linear isometric mapping  
        \begin{eqnarray}
                I:  \quad    {{\mathbb{C}}^{2j+1}}        \longrightarrow  {\mathrm L}^2({\mathbb{C}})   \;, \quad
                             \ket \psi  \longmapsto      \psi \; ,
        \end{eqnarray}
where the function $\psi$ on ${\mathbb{C}} \cong \mathbb R \times \mathbb R$ is defined
by its values $\psi(z) := \braket{z}{\psi}$.
        
The (Hilbert) adjoint $I^\dagger$ of $I$ explicitly reads
        \begin{eqnarray} 
                I^\dagger: \quad {\mathrm L}^2({\mathbb{C}})   \longrightarrow  {{\mathbb{C}}^{2j+1}}    \;, \quad
                               \varphi  \longmapsto     
                                         \int_{\mathbb{C}} \! d^2z \, \varphi(z) \ket{z} \, ,
        \end{eqnarray}
and the isometric property is simply stated as $I^\dagger I = \one$. 
The orthogonal projection from ${\mathrm L}^2({\mathbb{C}})$ onto
$I({{\mathbb{C}}^{2j+1}})$ is the operator $I I^\dagger =: E_0$.

Every operator $\cal B$ on ${{\mathbb{C}}^{2j+1}}$ can be realized by the unitary equivalent 
$I {\cal B} I^\dagger$ on 
$E_0({\mathrm L}^2({\mathbb{C}}))=I({\mathbb{C}}^{2j+1})$, which
trivially extends to the whole of ${\mathrm L}^2({\mathbb{C}})$.
In particular, it follows from (\ref{Hdef}) that
        \begin{equation} \label{ihiehe}
                I {\mathcal H} I^\dagger = E_0  H  E_0 \; ,
        \end{equation}
where $H$ is the bounded multiplication operator on ${\mathrm L}^2({\mathbb{C}})$ defined by
the function $h$, that is,
$ (H \varphi)(z) := h(z) \varphi(z)$ for all
$\varphi \in {\mathrm L}^2({\mathbb{C}})$. 
Furthermore, the embedded operator $I {\mathcal H} I^\dagger$ possesses a
continuous 
integral kernel $I {\mathcal H} I^\dagger (z, z')$ 
(also known as its position representation)
given by the coherent representation of ${\mathcal H}$, that is,
        \begin{equation} \label{intkern}
                I {\mathcal H} I^\dagger (z, z') = \bra{z} {\mathcal H} \ket{z'} \, .
        \end{equation}
%
Using \eq{ihiehe}, one can now verify the identity
$
        I \ep{-t {\mathcal H}} I^\dagger = E_0 \ep{-t E_0 H E_0} 
$
to all orders in $t$, which, analogous to \eq{intkern}, shows that
$E_0 \ep{-t E_0 H E_0}$ has a continuous integral kernel given by
the equation
\begin{equation} \label{intkernEhE}
                E_0 \ep{-t E_0 H E_0}(z,z') = 
                \bra{z} \ep{-t{\mathcal H}} \ket{z'} \, .
        \end{equation}
%
%
\subsection{A Schr{\"o}dinger operator and its ground-state eigenspace}\label{schroedi}
%
%
Consider on ${\mathrm L}^2({\mathbb{C}})$ the ``magnetic'' Schr\"odinger operator,
        \begin{equation} \label{reg}
                R := (i \partial_1 + A_1)^2 +
                        (i \partial_2 + A_2)^2 +  V
        \end{equation}
with the partial differential operators
$\partial_1 := \textstyle \partial / \textstyle \partial z_1$, 
$\partial_2 := \textstyle \partial / \textstyle \partial z_2$ and
the vector and scalar potentials $\left({A_1 \atop A_2}\right)$ and
$V$ acting as multiplication operators 
defined by the bounded and continuous functions, 
        \begin{eqnarray} \label{vec}
                \left({a_1(z)}\atop {a_2(z)}\right) &:=&
                                \left({-\partial_2 \atop \partial_1} \right) 
                         \ln g(z) = \frac{2(j+1)}{1+\abs{z}^2}
                         \left({ z_2 \atop  -z_1} \right) \, , \\
                v(z) &:=& \partial_1 a_2(z) - \partial_2 a_1(z) 
                                        = - \frac{4(j+1)}{(1+\abs{z}^2)^2} \, . \label{pot}
        \end{eqnarray}
The self-adjoint operator $R$
is tailored such that its ground-state eigenspace is identical to $E_0({\mathrm L}^2({\mathbb{C}}))$ and
the corresponding eigenvalue vanishes. In essence, this
follows from a result 
of Aharonov and Casher \cite{AhCa79} on zero-energy eigenstates. 
Since the proof is quite short, we will give it, thereby closely 
following the presentation in Ref. \onlinecite{CFKS87}.
%
%
Factorized like $R = D^\dagger D $, where 
$D := i \partial_1 + \partial_2 + A_1 - i A_2$,
the positivity of $R$
becomes manifest. Its null space consists of all those functions $\psi$ in 
${\mathrm L}^2({\mathbb{C}})$ with $D \psi = 0$. The general solution of
this differential equation is a product 
$\psi = g \phi$, where $\phi$ is any function analytic in $z^*$, 
that is, $(\partial_1-i \partial_2)\phi = 0$. Due to (\ref{weight}),
square integrability then requires $\phi$ to be any polynomial in $z^*$ 
of maximal degree $2j$, which proves that the ground-state eigenspace of $R$
and the subspace $E_0({\mathrm L}^2({\mathbb{C}}))= I({{\mathbb{C}}^{2j+1}})$ are identical.

%
%
Two remarks are in order.
\begin{enumerate}
\item
The spectrum of $R$ coincides with the  positive half-line, 
as can be inferred from Theorem 6.1 in  Ref.  \onlinecite{CFKS87}.
Following arguments as in the proof of Theorem 6.2 in Ref. \onlinecite{CFKS87},
one sees that zero is the 
only eigenvalue. 
Therefore the nature of the spectrum and the ground-state
eigenfunctions are explicitly known.  
However, we are not aware of explicit results on generalized eigenfunctions 
corresponding to strictly positive spectral values.
\item
Employing the spectral theorem, one proves that the semigroup  
generated by $\nu R$ converges strongly to the ground-state projection
$E_0$, in the sense that
        \begin{equation} \label{pro}
          \lim_{\nu \to  \infty} \norm{\ep{- t \nu R} \varphi -
        E_0 \varphi } = 0 \, ,\qquad 
        \mbox{for all $\varphi \in {\mathrm L}^2({\mathbb{C}})$ and $t > 0$,}
        \end{equation}
where the norm $\norm{\bcdot} := (\bcdot{\small|}\bcdot)^{1/2}$ corresponds to the
standard scalar product on ${\mathrm L}^2({\mathbb{C}})$.
\end{enumerate}
%
%
\subsection{The spin semigroup as the limit of a Schr{\"o}dinger semigroup}
%
%
With the material gathered in Secs.
\ref{einbetten} and \ref{schroedi} we can
isolate the central reason for the validity of  
the main result \eq{result} of the present paper. 
The point is that the spin semigroup,
now realized on ${\mathrm L}^2({\mathbb{C}})$,
can be understood as the limit $\nu \to \infty$ of the Schr\"odinger semigroup generated by
$\nu R + H$.
More precisely, we will show that the continuous integral kernel given in \eq{intkernEhE} is
the pointwise limit 
        \begin{equation} \label{limkern}
                E_0 \ep{-t E_0 H E_0}(z,z') =
                \lim_{\nu \to \infty} 
                \ep{-t (\nu R + H)}(z,z')\, , \qquad \mbox{for all $z, z' \in {\mathbb{C}}$ and $t>0$}, 
        \end{equation}
where the prelimit expression is the continuous integral kernel of $\exp\{-t(\nu R + H)\}$.
By expressing this integral kernel in terms of
the {\it Feynman-Kac-It\^o formula} 
\cite{Sim79Thm,BHL98} (observing $\partial_1 a_1 + \partial_2 a_2 =0$)
        \begin{eqnarray} \label{FKI}
                \ep{-t (\nu R + H)}(z,z') = & & 
                \dmut   \exp\left\{ (j+1) \int_0^t \! ds \,  
                \frac{\dot{b}(s) b^*(s) - \dot{b}^*(s) b(s)}{1+|b(s)|^2}\right\} \nonumber \\ 
                && \qquad \times \exp\left\{ 4(j+1)\nu \int_0^t \frac{ds}{(1 + |b(s)|^2)^2}
                -  \int_0^t \! ds \, h(b(s))\right\} \; , 
        \end{eqnarray}
the right-hand sides of \eq{limkern} and \eq{result} are seen to coincide. 

The proof of \eq{limkern} makes essential use of the semigroup property of $\ep{-t(\nu R + H)}$.
Throughout the proof we fix $t>0$ and pick some reference diffusion constant $\nu_0 >0$.
As a starting point we define
\begin{equation} \label{bring1}
     \eta_w^{(\lambda)}(z):= \ep{-t(\nu_0 R+\lambda H)}(z,w) \, ,
\end{equation}
for all $\lambda \in \mathbb R$ and $w,z \in {\mathbb{C}}$.
We assert that the function $\eta_w^{(\lambda)}: z \mapsto \eta_w^{(\lambda)}(z)$ is
continuous, bounded, and lies in ${\mathrm L}^2({\mathbb{C}})$. The continuity follows from that of the integral kernel in
\eq{FKI}. 
Boundedness and square-integrability result from the inequality
\begin{equation}
\abs{\eta_w^{(\lambda)}(z)} \le \ep{4(j+1)t \nu_0 } \, \ep{t \abs{\lambda} \norm{h}_\infty}
                             \, (4 \pi t \nu_0 )^{-1} \, \ep{-\abs{z-w}^2/4t \nu_0  }\;,
\end{equation}
where $\norm{h}_\infty := \sup_{z \in {\mathbb{C}}} \abs{h(z)} < \infty$
denotes the supremum norm of $h$. This inequality, in turn,
is found by estimating the path integral in \eq{FKI}.
We also state that the mappings $ w \mapsto \eta_w^{(\lambda)}$ and $\lambda \mapsto \eta_w^{(\lambda)}$
are strongly continuous. The first statement holds because of 
$
   \bigl({\displaystyle \eta_{w}^{(\lambda)}|\eta_{w'}^{(\lambda)}}\bigr) 
= \ep{-2t(\nu_0 R+\lambda H)}(w,w')
$ and the continuity of the integral kernel. The second one is a consequence of the inequality
\begin{equation}
  \norm{\eta_w^{(\lambda)} - \eta_w^{(\lambda')}} \le
      \sqrt{\frac{t }{8 \pi \nu_0}} \abs{\lambda - \lambda'} \norm{h}_\infty \, \ep{4(j+1)t \nu_0 } \, 
                        \ep{t \max\{\abs{\lambda},\abs{\lambda'}\} \norm{h}_\infty}
  \; ,
\end{equation}
which is derived by estimating the difference of two path integrals 
of type \eq{FKI} using the elementary inequality
$\abs{\ep x - \ep y } \le \abs{ x - y} \ep{\max\{x,y\}}$, for $ x,y \in \mathbb R$ .

The following two steps of the proof are based on writing the integral kernel 
for $\nu > 2 \nu_0$
as a scalar product,
\begin{equation} \label{semi} 
    \ep{-t(\nu R + H)}(z,z') 
 =  \left( \eta_z^{(\nu_0/\nu)} \bigl| \ep{- t(\nu-2\nu_0)(R + H/\nu)} \eta_{z'}^{(\nu_0/\nu)}\bigr. \right) \; .
\end{equation}
In the first step, we claim that
\begin{equation} \label{cl1}
 \lim_{\nu \to \infty} 
   \left( \eta_z^{(\nu_0/\nu)} \bigl| \ep{- t(\nu-2\nu_0)(R + H/\nu)} \eta_{z'}^{(\nu_0/\nu)} \bigr.\right)
     = \left( \eta_z^{(0)} \bigl| E_0 \ep{-t E_0 H E_0} \eta_{z'}^{(0)} \bigr.\right) \, ,
\end{equation}
for all $z, z' \in {\mathbb{C}}$.

Due to the strong continuity of $\lambda \mapsto \eta_w^{(\lambda)}$,
the boundedness of $\ep{- t(\nu-2\nu_0)(R + H/\nu)}$,
which is uniform in $\nu$, and the continuity of the scalar product $(\bcdot | \bcdot)$, 
it suffices to show that
\begin{equation} \label{st-op} 
        \lim_{\nu \to \infty} \norm{ \ep{- t(\nu-2\nu_0)(R + H/\nu)} \varphi  - E_0 \ep{-t E_0 H E_0} \varphi }= 0 \qquad 
                                                     \mbox{for all $\varphi \in {\mathrm L}^2({\mathbb{C}})$} \, . 
\end{equation}   
To prove this strong operator convergence we employ the Duhamel-Dyson-Phillips 
perturbation expansion,
\begin{eqnarray}
  \ep{-t(\nu-2\nu_0)(R+ H/\nu)} \varphi & = & \ep{-t(\nu-2\nu_0)R}\varphi + 
                \sum_{n=1}^\infty  \left({\frac{2\nu_0 - \nu}{\nu}}\right)^n \nonumber \\
                && \times \int_0^t\! \mbox{\it ds}_n 
\cdots
                \int_0^{s_3}\!\mbox{\it ds}_2 \int_0^{s_2}\!\mbox{\it ds}_1 \,
                \ep{-(t-s_n)(\nu-2\nu_0)R}\, H \nonumber \\
                && \qquad \times \cdots \times \ep{- (s_2-s_1)(\nu-2\nu_0) R} H \ep{- s_1 (\nu-2\nu_0) R} \varphi  \, ,
\end{eqnarray}
which
converges uniformly in $\nu \in ]2\nu_0, \infty[$ with respect to 
the norm on ${\mathrm L}^2({\mathbb{C}})$. This holds because the norm of the series
is dominated by the exponential series  
$
        \sum_{n=0}^\infty (t^n/n!) \norm{h}_\infty^n \norm{\varphi}
$, independent of $\nu$. 
Thus, the limit $\nu \to \infty$ can be 
interchanged with the summation and, using (\ref{pro})
and the dominated-convergence theorem,
we obtain the expansion of $E_0 \ep{-t E_0 H E_0}\varphi$.

In the second and final step we claim that the right-hand side of \eq{cl1} 
is already the desired
integral kernel, that is,
\begin{equation}
   \left( \eta_z^{(0)} | E_0 \ep{-t E_0 H E_0} \eta_{z'}^{(0)} \right) 
     = E_0 \ep{-t E_0 H E_0} (z,z') \, , \qquad \mbox{for all $z,z' \in {\mathbb{C}}$} \, .
\end{equation}
This is verified by checking that the mapping
$
        (z,z') \mapsto  ( \eta_z^{(0)} | E_0 \ep{-t E_0 H E_0} \eta_{z'}^{(0)} )
$
constitutes an integral kernel of $E_0 \ep{-t E_0 H E_0}$
and is, in fact, continuous.  
The former is true since
$\ep{-t\nu_0 R} E_0 = E_0$.
The latter holds because the mapping $w \mapsto \eta_w^{(0)}$ is strongly continuous,
the operator $E_0 \ep{-t E_0 H E_0}$ is bounded, and the scalar product $(\bcdot{\small|}\bcdot)$ is continuous.

%
%

\section{Concluding remarks}
We conclude the paper with six remarks.
\begin{enumerate}
\item
As already mentioned in Sec. \ref{defres}, the main result \eq{result} 
cannot  be obtained from a result in Ref. \onlinecite{DaKl85} merely by
stereographically projecting the Brownian paths from the two-sphere $S^2$ onto
the (extended) Euclidean plane ${\mathbb{R}}^2$. The reason can be traced back to the different 
operators, or equivalently path measures, 
used for regularization. The stereographic 
projection corresponds to re-expressing the differential operator on ${\mathrm L}^2(S^2)$
used by the authors of Ref. \onlinecite{DaKl85} in flat Cartesian co-ordinates.
The resulting operator is not of the standard Schr\"odinger form, acts on a
weighted Hilbert space, and is not related to planar Brownian motion.
\item 
In contrast to Ref. \onlinecite{DaKl85} the regularizing operator $R$ used 
in the proof of \eq{limkern}, and hence of 
\eq{result}, has no spectral gap above its ground-state eigenvalue. 
Accordingly, $\ep{-t \nu R}$ only converges strongly, and not in operator norm,
to the corresponding eigenprojection $E_0$ as $\nu \to \infty$. 
As a consequence, the foregoing proof of the pointwise convergence of integral kernels 
required a strategy  different 
from that in Ref. \onlinecite{DaKl85}.
\item
From a fundamental point of view, it is gratifying that a spin system 
can be related to a limit of a well-defined integral over continuous 
Brownian-motion paths.
From a practical point of view, it would be desirable to apply to \eq{result}
the well-established theory and computational possibilities
associated with the 
flat-space Wiener measure,\cite{Sim79,ReYo94,ESY93} in order to attack 
specific spin problems of physical interest.
One such problem, which has been extensively discussed in the recent
literature,
\cite{Kur92,FKSF95,FKNS95,Koc95,EMNP96,ShTa98} is to understand the nature
of the saddle-point approximation for the evaluation of
continuum path integrals connected with
simple spin Hamiltonians. Looking at Table \ref{tabelle} and the 
resulting $j$-dependence of the path integrand
in \eq{result}, this approximation is expected to be the more reliable the 
larger the spin quantum number is.
Moreover, for Hamiltonians ${\cal H}$ linear in the spin operators, the saddle-point
approximation is believed \cite{Kur92,FKSF95,FKNS95,Koc95,EMNP96}
to give the (explicitly known) exact result already 
for given finite $j$.
In this context, when dealing with symbolic continuum
path integrals one has to overcome the so-called overspecification problem
due to missing regularizing terms in the action functionals of those 
path integrals.\cite{Kla79,EMNP96}
Rigorous continuum path integrals as used in \eq{result}
do not suffer from this problem by their very construction. 
Of course, the details for the 
saddle-point approximation 
of the Wiener type of path integral in the ultradiffusive limit
still have to be worked out. 
\item
In Refs. \onlinecite{Mar92} and \onlinecite{AKL93} the
ground-state eigenspace of a charged point mass
under the 
influence of a certain magnetic field on an even-dimensional
Riemannian manifold is studied, thereby extending the 
Aharonov-Casher theorem. \cite{AhCa79,CFKS87} This result lies at the heart of the
quantization procedure 
proposed in Refs. \onlinecite{AKL93,AlKl96,KlOf89}.
A quantum system is hereby represented on the ground-state eigenspace of such a 
generalized Landau Hamiltonian on the Hilbert space of functions over its
classical phase space. 
The symplectic structure of the latter determines the magnetic field.
In this sense, Eq. \eq{result} read from right to left can be viewed as a
{\it quantization prescription} for a classical spin system.

In this context it is worth mentioning that the path integral
in \eq{result} is well-defined for all values of $j$ taken from the positive half-line.
Even more, in the limit $\nu \to \infty$ it manages to single out the set of allowed
spin quantum numbers, $\{0,1/2,1,3/2,\dots\}$, from the ``classical'' continuum $[0,\infty[$.\\
More precisely, for a given bounded and continuous
$h: {\mathbb C} \rightarrow {\mathbb C}$
and $j \in [0,\infty[$ we assert that the right-hand side
of \eq{result} 
is equal to $\bra{\psi(z)}\ep{-t{\mathcal H}_\psi}\ket{\psi(z')}$. Here the set of vectors,
\begin{equation}
  \ket{\psi(z)} := g(z) \sum_{n=0}^{2(j)} \sqrt{\left({2j \atop n}\right)} z^n \ket{\psi_n},
\qquad z \in \mathbb C \, ,
\end{equation}
is unity-resolving in ${\mathbb C}^{2(j)+1}$,
where $(j)$ denotes the smallest integer
or half-integer equal to or larger than $j$, and $\{\ket{\psi_n}\}$ is
a fixed but arbitrary
orthonormal basis in ${\mathbb C}^{2(j)+1}$. The binomial coefficient can be defined
recursively by $\bigl({2j \atop 0}\bigr):=1$ and
$\bigl({2j \atop n+1}\bigr) := \frac{2j-n}{n+1} \bigl({2j \atop n}\bigr)$,
and $g(z)$ is defined by \eq{weight} as it stands for general $j \in [0,\infty[$.
Finally, ${\mathcal H}_\psi$ is an operator on ${\mathbb C}^{2(j)+1}$ associated to the given $h$
by the definition
\begin{equation}
  {\mathcal H}_\psi := \int_{\mathbb C} d^2z \, h(z) \ket{\psi(z)}  \bra{\psi(z)} \; .
\end{equation}
This association can be viewed as a quantization, which maps the pair $(j,h)$
to the pair $\left( (j) , {\mathcal H}_\psi\right)$ with
${\mathcal H}_\psi$ being interpreted
as the Hamiltonian of a spin with quantum number $(j)$.
While ${\mathcal H}_\psi$ in general depends on the chosen basis
$\{\ket{\psi_n}\}$, 
the expression $\bra{\psi(z)} \ep{-t {\mathcal H}_\psi} \ket{\psi(z')}$
does not because of unitary invariance.

For the proof of the above assertion we remark that the latter is identical to \eq{result}
in the case $j=(j)$, because then $\ket{\psi(z)} = \ket{z}$ when choosing $\ket{\psi_n} = \ket{j,n-j}$,
the usual orthonormal eigenbasis of ${\mathcal J}_3$. In the case $j<(j)$,
the proof follows from \eq{FKI},
equations analogous to \eq{limkern} and \eq{intkernEhE}, and the Aharonov-Casher theorem,
which in our setting states that the ground-state eigenspace of the ``magnetic'' Sch\"odinger operator $R$
(stemming from $g$, confer \eq{reg} -- \eq{pot}) has a dimension equal to the largest
integer strictly smaller than
$|\int_{\mathbb C} d^2z \, v(z) |/ 2\pi =2j+2$
and is spanned by the set of orthonormal functions $z \mapsto \braket{\psi(z)}{\psi_n}$,
$n=0,1,\dots, 2(j)$.
\item
It is straightforward to generalize formula \eq{result} to systems where the
Hamiltonian $\cal H$ depends explicitly on time and/or several (coupled) spins.
The formula in the latter case, like its older ``spherical relative'' in
Ref. \onlinecite{DaKl85}, may then serve as a rigorous starting point
for the derivation
of effective field theories, which aim to descibe the low-energy excitations
of quantum lattice models for magnetism. Confer, for example, Refs.
\onlinecite{FrSt88,Fra91,Aue98}, and references therein.
\item
Following the reasoning of the present paper it should also be 
straightforward to derive flat-space Wiener-regularized 
path integrals for physical systems with 
degrees of freedom that are neither
of the canonical nor of the spin type. 
\end{enumerate}
\section*{Acknowledgments}
The authors are grateful to G. Junker, J. R. Klauder, and E. A. Kochetov
for stimulating discussions and helpful remarks. One of the authors (B. B.) 
would like to thank 
the Institut f\"ur Theoretische Physik at the Universit\"at Erlangen-N\"urnberg
for the kind hospitality during the final stage of the preparation of
this work.

%



\begin{references}
%
%
%
\bibitem{Fey48}{R. P. Feynman, {\it Space-time approach to non-relativistic quantum
mechanics}, Rev. Mod. Phys. {\bf 20}, 367-387 (1948)}
%
\bibitem{FeHi65}{R. P. Feynman, A. R. Hibbs, {\it Quantum mechanics and path integrals}, McGraw-Hill, New York 1965}
%
\bibitem{Sim79}{B. Simon, {\it Functional integration and quantum
physics}, Academic, New York 1979}
%
\bibitem{Sch80}{L. S. Schulman, {\it Techniques and applications of path integration}, corr.
reprint of the 1981 edition, Wiley, New York 1996}
%
\bibitem{Kle95}{H. Kleinert, {\it Path integrals in quantum mechanics, statistics,
and polymer physics}, 2nd edition, World Scientific, Singapore 1995}
%
\bibitem{Roe94}{G. Roepstorff, {\it Path integral approach to quantum physics}, Springer, Berlin 1994}
%
\bibitem{Mar59a}{J. L. Martin, {\it Generalized classical dynamics, and the 'classical analogue' of a Fermi oscillator}, Proc. Roy. Soc. London A {\bf 251}, 536-542 (1959)}
%
\bibitem{Mar59b}{J. L. Martin, {\it The Feynman principle for a Fermi system}, Proc. Roy. Soc. London A {\bf 251}, 543-549 (1959)}
%
\bibitem{Kla60}{J. R. Klauder, {\it The action option and a Feynman quantization of spinor fields
in terms of ordinary c-numbers}, Ann. Phys. (N. Y.) {\bf 11}, 123-168 (1960)}
%
%
\bibitem{Kla79}{J. R. Klauder, {\it Path integrals and stationary-phase approximations}, Phys.\ Rev.\ D {\bf 19}, 2349-2356 (1979)}
%
\bibitem{KuSu80}{H. Kuratsuji, T. Suzuki, {\it Path integral in the representation of $SU(2)$ coherent state and classical
                 dynamics in a generalized phase space}, J. Math. Phys. {\bf 21}, 472-476 (1980)} 
%
\bibitem{DaKl85}{I. Daubechies, J. R. Klauder, {\it
Quantum-mechanical path integrals with Wiener measure for all
polynomial Hamiltonians II}, J.\ Math. Phys. {\bf 26}, 2239-2256
(1985)}
%
\bibitem{Sch68}{L. Schulman, {\it A path integral for spin}, Phys. Rev. {\bf 176},
1558-1569 (1968)}
%
\bibitem{IJR98}{A. Inomata, G. Junker, C. R\"osch, {\it Remarks on the magnetic top},
Found. Phys. {\bf28}, 735-739 (1998)}
%
\bibitem{NiRo88}{H. \ B. Nielsen, D. Rohrlich, {\it A path integral to quantize spin}, Nucl. Phys. B {\bf 299}, 471-483 (1988)}
%
\bibitem{Joh89}{K. Johnson, {\it Functional integrals for spin}, Ann. Phys. (N. Y.) {\bf 192}, 104-118 (1989)}
%
\bibitem{FrSt88}{E. Fradkin, M. Stone, {\it Topological terms in one- and two-dimensional quantum Heisenberg antiferromagnets}, Phys.\ Rev.\ B {\bf 38}, 7215-7218 (1988)}
%
\bibitem{Sto89}{M. Stone, {\it Supersymmetry and the quantum mechanics of spin}, Nucl.\ Phys.\ B {\bf 314}, 557-586 (1989)}
%
\bibitem{JaKu91}{T. Jaroszewicz, P. S. Kurzepa, {\it Spin, statistics, and geometry of random walks}, Ann. Phys. (N. Y.) {\bf 210}, 255-322 (1991)}
%
\bibitem{CDGR97}{D. C. Cabra, A. Dobry, A. Greco, G. L. Rossini, {\it On the path integral representation for spin systems}, J. Phys. A {\bf 30}, 2699-2704 (1997)}
%
\bibitem{AnJo82}{G. F. De Angelis, G. Jona-Lasinio, {\it A stochastic description of a spin-$1/2$ particle in a magnetic field}, J.\ Phys.\ A {\bf 15}, 2053-2061 (1982)}
%
\bibitem{Lem96}{L.\ F.\ Lemmens, {\it The Ehrenfest model: A path-integral for spin}, Phys. Lett. A {\bf 222}, 419-423 (1996)}
%
\bibitem{Kur92}{H. Kuratsuji, {\it Path integrals in the $SU(2)$ coherent state representation and related topics}, in:
A.\ Inomata, H.\ Kuratsuji, C.\ C.\ Gerry, {\it Path integrals and coherent states of $SU(2)$ and $SU(1,1)$},
                World Scientific, Singapore 1992, pp.\ 139-218}
%
\bibitem{FKSF95}{K. Funahashi, T. Kashiwa, S. Sakoda, K. Fujii,
{\it Coherent states, path integral, and semiclassical approximation}, J. Math. Phys. 
{\bf 36}, 3232-3253 (1995)}
%
\bibitem{FKNS95}{K. Funahashi, T. Kashiwa, S. Nima, S. Sakoda,
{\it More about path integrals for spin}, Nucl. Phys. B {\bf 453},
508-528 (1995)}
%
\bibitem{Koc95}{E.\ A.\ Kochetov, {\it SU(2) coherent-state path integral}, J.\ Math.\ Phys.\ {\bf 36}, 4667-4679 (1995)}
%
\bibitem{EMNP96}{E. Ercolessi, G. Morandi, F. Napoli, P. Pieri, 
{\it Path integrals for spinning particles, stationary phase and the Duistermaat-Heckman theorem},
J. Math. Phys. {\bf 37}, 535-553 (1996)}
%
\bibitem{ShTa98}{J. Shibata, S. Takagi, {\it A note on (spin-)coherent-state path integral}, quant-ph/9807005}
%
\bibitem{Rad71}{J.\ M.\ Radcliffe, {\it Some properties of coherent spin states}, J.\ Phys.\ A {\bf 4}, 313-323 (1971)}
%
\bibitem{KlSk85}{J. R. Klauder and B.-S. Skagerstam,
{\it Coherent states -- {\it A}pplications in physics and
                  mathematical physics}, World Scientific, Singapore 1985}
%
\bibitem{Per86}{A.\ Perelomov, {\it Generalized coherent states and
their applications}, Springer, Berlin 1986}
%
%
\bibitem{ZFG90}{W.-M.\ Zhang, D.\ H.\ Feng, R.\ Gilmore, {\it Coherent states: Theory and some applications}, Rev.\ Mod.\ Phys.\
    {\bf 62}, 867-927 (1990)}
%
\bibitem{Ber72a}{F. A. Berezin, {\it Covariant and contravariant symbols of operators},
        Math.\ USSR Izvestija \ {\bf 6}, 1117-1151 (1972),
russ. orig.: Izv. Akad. Nauk SSSR, Ser. Mat. {\bf 36}, 1134-1167
(1972)}
%
\bibitem{Sim80}{B. Simon, {\it The classical limit of quantum partition functions}, 
Commun. Math. Phys. 
{\bf 71 }, 247-276 (1980)}
%
\bibitem{Kla98}{J. R. Klauder, {\it Coherent state path integrals at (nearly) 40},
in: R. Casalbuoni, R. Giachetti, V. Tognetti, R. Vaia, P. Verrucchi (eds.), 
{\it Path integrals from peV to TeV: 50 years after Feynman's paper}, World Scientific, Singapore 1999, pp.\ 65-70, also available at:  quant-ph/9810043}
%
\bibitem{Lie73}{E.\ H.\ Lieb, {\it The classical limit of quantum spin systems}, Commun. Math. Phys. {\bf 31}, 327-340 (1973)}
%
%
\bibitem{RoWi87}{L. C. G. Rogers, D. Williams, {\it Diffusions, Markov processes, and martingales}, vol. 2: It\^o calculus, Wiley, Chichester 1987}
%
\bibitem{Pro95}{Ph. Protter, {\it Stochastic integration and differential equations - A new approach},
        3rd printing, Springer, Berlin 1995}
%
\bibitem{ReYo94}{D. Revuz, M. Yor,
{\it Continuous martingales and {B}rownian motion}, 3rd edition,
Springer, Berlin 1999}
%
\bibitem{RoWiTheorem}{Theorem IV.46.4 in \ Ref. \ \onlinecite{RoWi87}}
%
\bibitem{ReYoTheorem}{Exercise IV.2.18 in \ Ref. \ \onlinecite{ReYo94}}
%
\bibitem{Ber74a}{F. A. Berezin,
         {\it Quantization},
         Math. USSR Izvestija {\bf 8}, 1109-1165 (1974),
         russ. orig.: Izv. Akad. Nauk SSSR, Ser. Mat. {\bf
                  38}, 1116-1175 (1974)}
%
\bibitem{Ber74b}{F. A. Berezin,
         {\it General concept of quantization}, Commun. Math. Phys. {\bf 40}, 153-174 (1975)}
%
\bibitem{ACGT72}{F.\ T.\ Arecchi, E. Courtens, R. Gilmore, H. Thomas, {\it Atomic coherent states in quantum optics},
                 Phys.\ Rev.\ A {\bf 6}, 2211-2237 (1972)}
%
\bibitem{Gil74}{R. Gilmore, {\it Lie groups, Lie algebras, and some of their applications}, Wiley, New York 1974}
%
\bibitem{DaKl86}{I. Daubechies, J. R. Klauder, {\it True measures for real time path integrals},
  in:    M. C. Gutzwiller, A. Inomata, J. R. Klauder, L. Streit (eds.),
{\it Path integrals from me{V} to {M}e{V}},
World Scientific, Singapore 1986, pp. 425-432}            
%
\bibitem{AhCa79}
{Y. Aharonov, A. Casher, {\it Ground state of spin-$1/2$ charged
particle in a two-dimensional magnetic field}, Phys.\ Rev.\  A
{\bf19}, 2461-2462 (1979)}
%
%
\bibitem{CFKS87}{H. L. Cycon, R. G. Froese, W. Kirsch, B. Simon,
{\it Schr\"odinger operators}, Springer, Berlin 1987}
%
\bibitem{Sim79Thm}{Theorems 14.5 and 15.5 in Ref. \onlinecite{Sim79}}
\bibitem{BHL98}{K. Broderix, D. Hundertmark, H. Leschke, {\it Continuity properties of
           Schr\"odinger semigroups with magnetic fields}, 
        math-ph/9808004 and mp\_arc 98-564, to appear in Rev. Math. Phys.}
%
%
%
%
\bibitem{ESY93}{A. D. Egorov, P. I. Sobolevsky, L. A. Yanovich,
{\it Functional integrals: Approximate evaluation and applications}, Kluwer,
Dordrecht 1993}
%
\bibitem{Mar92}{P. Maraner, {\it Landau ground state on Riemannian surfaces},
Mod. Phys. Lett. A {\bf 7}, 2555-2558 (1992)}
%
\bibitem{AKL93}{R. Alicki, J. R. Klauder, J. Lewandowski, {\it Landau-level ground state and its relation for a general quantization procedure}, Phys. Rev. A {\bf 48}, 2538-2548 (1993)}
%
\bibitem{AlKl96}{R. Alicki, J.R. Klauder, {\it Quantization of systems with a general phase space equipped with a Riemannian metric}, J. Phys. A {\bf 29}, 2475-2483 (1996)}
%
\bibitem{KlOf89}{J. R. Klauder, E. Onofri, {\it Landau levels and geometric quantization}, Int. J. Mod. Phys. {\bf 4}, 3939-3949 (1989)}
%
%
\bibitem{Fra91}{E. Fradkin, {\it Field theories of condensed matter systems},
Addison-Wesley, Redwood City, CA 1991}
%
\bibitem{Aue98}{A. Auerbach, {\it Interacting electrons and quantum magnetism},
corr. 2nd printing, Springer, New York 1998}
\end{references}
\end{document}